\documentclass{webofc}
\usepackage[varg]{txfonts}   
\usepackage{soul}
\usepackage{dsfont}
\usepackage{subcaption}
\begin{document}
\title{Thermal dilepton production in heavy-ion collisions at beam-energy-scan (BES) energies}
%
%

\author{\firstname{Jessica} \lastname{Churchill}\inst{1}\fnsep\thanks{\email{jessica.churchill@mail.mcgill.ca}}, 
        \firstname{Lipei} \lastname{Du}\inst{1}\fnsep\thanks{\email{lipei.du@mail.mcgill.ca}},
        \firstname{Bailey} \lastname{Forster}\inst{1}\fnsep\thanks{\email{bailey.forster@mail.mcgill.ca}},
        \firstname{Han} \lastname{Gao}\inst{1}\fnsep\thanks{\email{han.gao3@mail.mcgill.ca}}, 
        \firstname{Greg} \lastname{Jackson}\inst{2}\fnsep\thanks{\email{jackson@subatech.in2p3.fr}},
        \firstname{Sangyong} \lastname{Jeon}\inst{1}\fnsep\thanks{\email{sangyong.jeon@mcgill.ca}},
        \firstname{Charles} \lastname{Gale}\inst{1}\fnsep\thanks{Speaker, \email{charles.gale@mcgill.ca}}
}

\institute{Department of Physics, McGill University, 3600 University street, Montreal, QC, Canada H3A 2T8
\and
           SUBATECH,  Universit\'e de Nantes, IMT Atlantique, IN2P3/CNRS,  4 rue Alfred Kastler, La Chantrerie BP 20722, 44307 Nantes, France
          }

\abstract{%
The lepton pair production rate at finite temperature and at next-to-leading-order (NLO) is calculated for quark-gluon plasma at non-zero baryon density. Yields are obtained using a (3+1)D multicomponent simulation   capable of  reproducing hadronic observables measured in the RHIC Beam Energy Scan. Spectra of intermediate invariant mass dileptons are compared with measurements from the STAR collaboration. The result of a study  where the temperature  information obtained from the dilepton spectrum from heavy-ion collisions performed at different energies and centralities is presented. 
}
\maketitle
\section{Introduction}
\label{intro}
High energy heavy-ion collisions  create a quark-gluon plasma (QGP), an exotic phase of strongly-interacting matter where quarks and gluons are the relevant degrees of freedom.  The bulk of current analyses of the collision remnants involve  hadrons, and the study of their collective motion has highlighted the success of viscous relativistic fluid dynamics as a modeling paradigm and as means to access QCD transport parameters \cite{Gale:2013da}. However, the conditions that influence strongly interacting particles are those that prevail close to hadronization and the surface of last scattering. Electromagnetic radiation, on the other hand, escapes the medium once emitted and can therefore report on  local conditions.  This is true for both real and virtual photons, and those probes are complementary. The real photon spectra will depend upon the choice of reference frames, and this Lorentz-variance can be used to guide the modeling of the collisions. The dilepton production rate is suppressed with respect to that of real photons by one power of $\alpha_{\rm em}$, but the invariant mass spectrum is independent of the local flow conditions. This work highlights results on lepton pair thermal production at RHIC Beam Energy Scan (BES) energies, using rates derived at NLO in the strong coupling and at non-vanishing net baryon chemical potential. 
\section{Theory: Dilepton production rates}
\label{Theory}
In a quark-gluon plasma, the differential rate $\Gamma_{\ell \bar{\ell}}$ of lepton pair production (per unit invariant mass and rapidity ) at finite temperature is related to the imaginary part of the retarded in-medium photon self-energy $\Pi^{\mu \nu}$ \cite{Weldon1990,Gale1990}:
\begin{eqnarray}
\frac{d \Gamma_{\ell \bar{\ell}}}{dM dy} = \frac{\alpha_{\rm em}^2}{3 \pi^3 M} \left(\sum_{i=1}^{n_f} Q_i^2 \right) \int d^2 {\bf k}_\bot \frac{{\rm Im}\, \Pi_\mu^\mu}{\exp\left(\beta \omega\right)-1},
\end{eqnarray}
where each of the light flavours ($n_f =3$) have charge fraction $Q_i$ (in units of the electron charge) and negligible bare mass. The lepton masses are also neglected from here on. The imaginary part of the self-energy defines a spectral function and  its transverse and longitudinal components:  $	{\rm Im}\, \Pi^{\mu \nu} = \rho^{\mu \nu} = {\mathds P}_{\rm L}^{\mu \nu}  \rho_{\rm L} + {\mathds P}_{\rm T}^{\mu \nu} \rho_{\rm T}$, where $\mathds{P^{\mu \nu}_{\rm T/L}}$ are projection operators.

\begin{figure}[h]
	\centering
	\includegraphics[width=2cm]{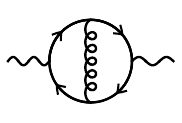}
	\includegraphics[width=2cm]{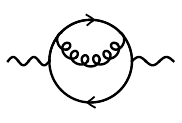}
	\includegraphics[width=2cm]{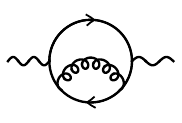}
    \raisebox{0.8ex}{\includegraphics[width=4cm]{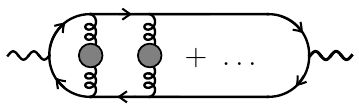}}
	\caption{The first three diagrams represent  strict two-loop contributions to the photon self-energy for $\mu_{\rm B} \ne 0$ and at NLO, and the last one is that of the LPM. The gluon lines with a shaded circle represent gluon propagators evaluated in the Hard Thermal Loops (HTL) limit  \cite{Kapusta:2023eix}.}
	\label{fig-1}       
\end{figure}

\begin{figure}[t!]
\begin{subfigure}{0.3\textwidth}
    \includegraphics[width=5cm]{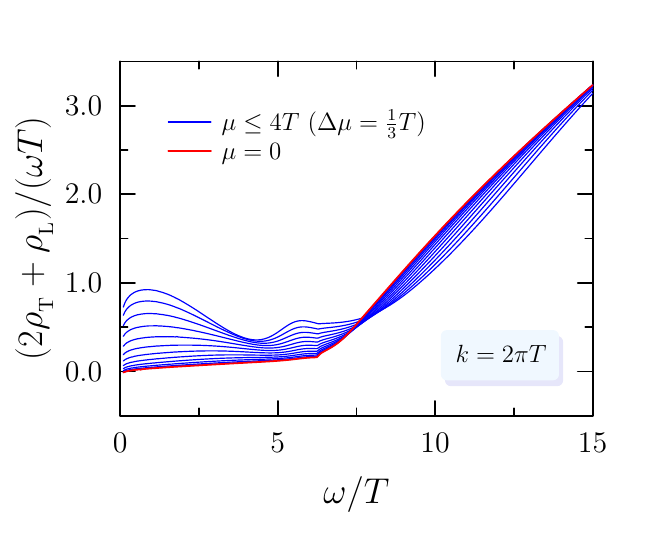}
\end{subfigure}
    \begin{subfigure}{0.3\textwidth}
   	\hspace*{0.8cm}
\includegraphics[width=5cm]{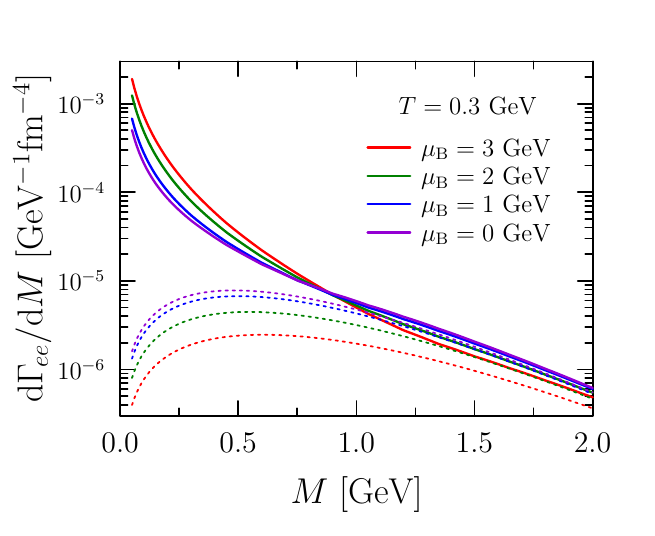}
\end{subfigure}
	\caption{
		The left panel shows $\rho^\mu_\mu = 2 \rho_{\rm T} + \rho_{\rm L}$, the trace of the  spectral density at NLO, as a function of the specific energy for a momentum of $k/T = 2 \pi$, for several values of the baryon chemical potential. The right panel shows the chemical potential dependence of the LO dilepton rates (dashed lines) and of the   rates which include NLO corrections (full lines). }
	\label{fig-2}       
\end{figure}

 In a fluid-dynamical environment, the production rate is integrated with a model which evolves the temperature ($T= 1/\beta$) and the four-volume. The invariant mass $M = \sqrt{\omega^2 - {\bf k}^2}$ is obtained from the energy $\omega$ and three-momentum ${\bf k}$ in the local rest frame, both of which will depend on the local flow velocity and on the rapidity $y$ \cite{Churchill:2023vpt}. In conditions when the net baryon chemical potential $\mu_{\rm B} \ne 0$, the Debye mass $m_D$ and the asymptotic quark mass $m_\infty$ are modified, such that  $m_D^2 \to  g^2 \left[\left( \frac{1}{2} n_f +N_c \right) \frac{T^2}{3} + n_f \frac{\mu^2}{2 \pi^2}\right]$, and  $m_\infty^2 \to \frac{g^2}{3} \left( T^2 + \frac{\mu^2}{\pi^2}\right)$, where $\mu = \mu_{\rm B}/3$.  The NLO corrections  appear at two-loop order and must also include the Landau-Pomeranchuk-Migdal (LPM) class of diagrams, as shown in Fig \ref{fig-1}. 
The combination of the explicit two-loop and the LPM diagrams has to be done with some care, as the LPM resummation is evaluated in the limit of  kinematics only strictly valid near the light cone. In order to access higher values of $M$ in the coherent addition of all sources  represented graphically in Fig. \ref{fig-1} , one must re-expand the LPM contribution to remove double-counting  \cite{Churchill:2023vpt,Ghisoiu2014}.  This is done at NLO and finite $\mu_{\rm B}$ here for the first time. The left panel of Fig. \ref{fig-2} shows the sum of the scaled spectral density  as a function of the scaled energy, for a value of the scaled momentum $k/T = 2 \pi $. The spectral density shows an enhancement with growing chemical potential in the deep spacelike region -- unaccessible to dilepton production experiments -- and suppression at higher energies in the timelike region. The right panel of Fig. \ref{fig-1} shows the invariant mass-dependent rate at a given temperature of $T=0.3$ GeV, for different values of the baryon chemical potential. The $\mu_{\rm B}$  dependence seen there is moderate compared with the large effects of NLO corrections at small invariant masses. However, the exploration of the different polarizations has the potential to reveal  more information about the baryonic content, as do more differential observables \cite{Bailey}. At low invariant masses, the effect of the NLO contribution is large and grows with $\mu$, owing to contributions of the bremsstrahlung-type absent at LO: this  is consistent with NLO estimates of real photon production \cite{Gervais2012}. As invariant mass grows so does the annihilation-type contribution present at LO, and known to be suppressed with increasing $\mu$ \cite{Dumitru1993}. 

\section{Dilepton yields}
\label{sec-2}
\subsection{Fluid dynamical modeling}
With the BES regime in mind, where collision energies range from $\sqrt{s_{\rm NN}} = $ 7.7 GeV to 200 GeV, dilepton rates are integrated to produce yields. The fluid dynamical modeling is carefully devised to reproduce hadronic features. This means that the initial states are first defined, the hydrodynamics equations of motion are then solved and the time-evolution performed. A hadronic afterburner evolves the late states to their kinetic freeze-out in a procedure described in detail in Ref. \cite{Churchill:2023vpt}. Note that the yields reported on here are not (yet) corrected for viscous effects even though the underlying hydrodynamical evolution is a viscous one \cite{Du:2023gnv}. Those corrections have been considered previously \cite{Vujanovic2013} and will be incorporated in a future study. 

\subsection{Extracting temperature information}
\label{subsect_2}
In the limit where $M \gg T$, the dilepton production rate in the frame where temperature is defined can be approximated as $d \Gamma_{\ell {\bar \ell}}/dM \sim (M T)^{3/2} \exp(-M/T)$. Assuming that the integrated spectrum follows a similar form, one can extract an effective temperature $T_{\rm eff}$ from the shape of the dilepton spectrum. In the ``low invariant mass region'', $0  < M \lesssim m_\phi$, it is know that in-medium hadronic reactions involving low mass mesons and baryons will contribute massively to the lepton pair thermal spectrum and will outshine the QGP signal, which however is predicted to dominate in the ``intermediate mass region'' ($m_\phi \lesssim M \lesssim m_{J/\psi}$)\cite{Rapp:2000pe}, provided that background sources such as Drell-Yan and the semi-leptonic decays of open charm mesons are  either small, or can be measured separately and subtracted. Consequently, this work reports on dilepton emission in the mass region $1 < M < 3$ GeV. 

\begin{figure}
	\sidecaption
	\includegraphics[width=4cm]{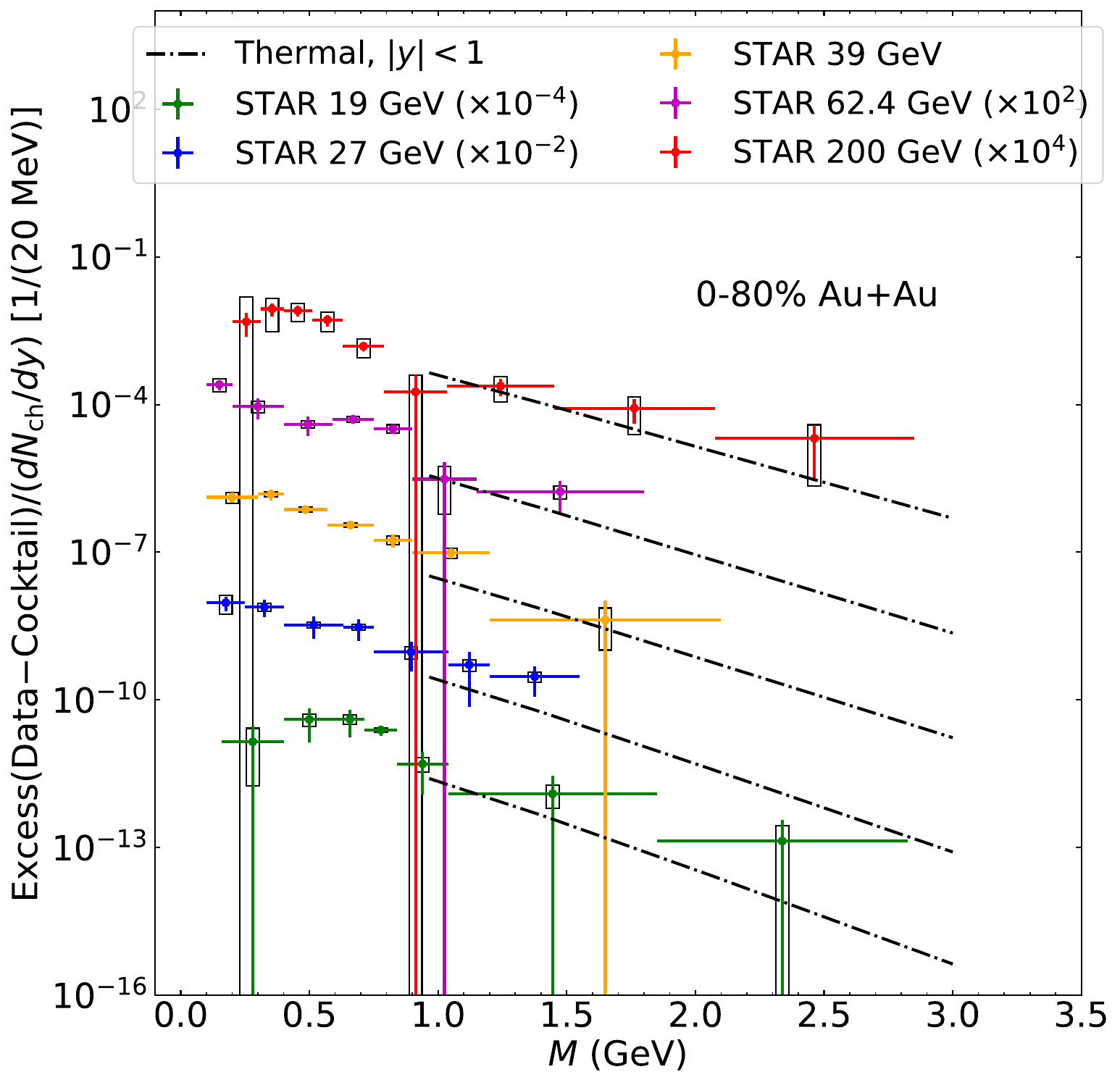}
\hspace*{0.3cm}\raisebox{-0.7ex}{\includegraphics[width=5.1cm,clip]{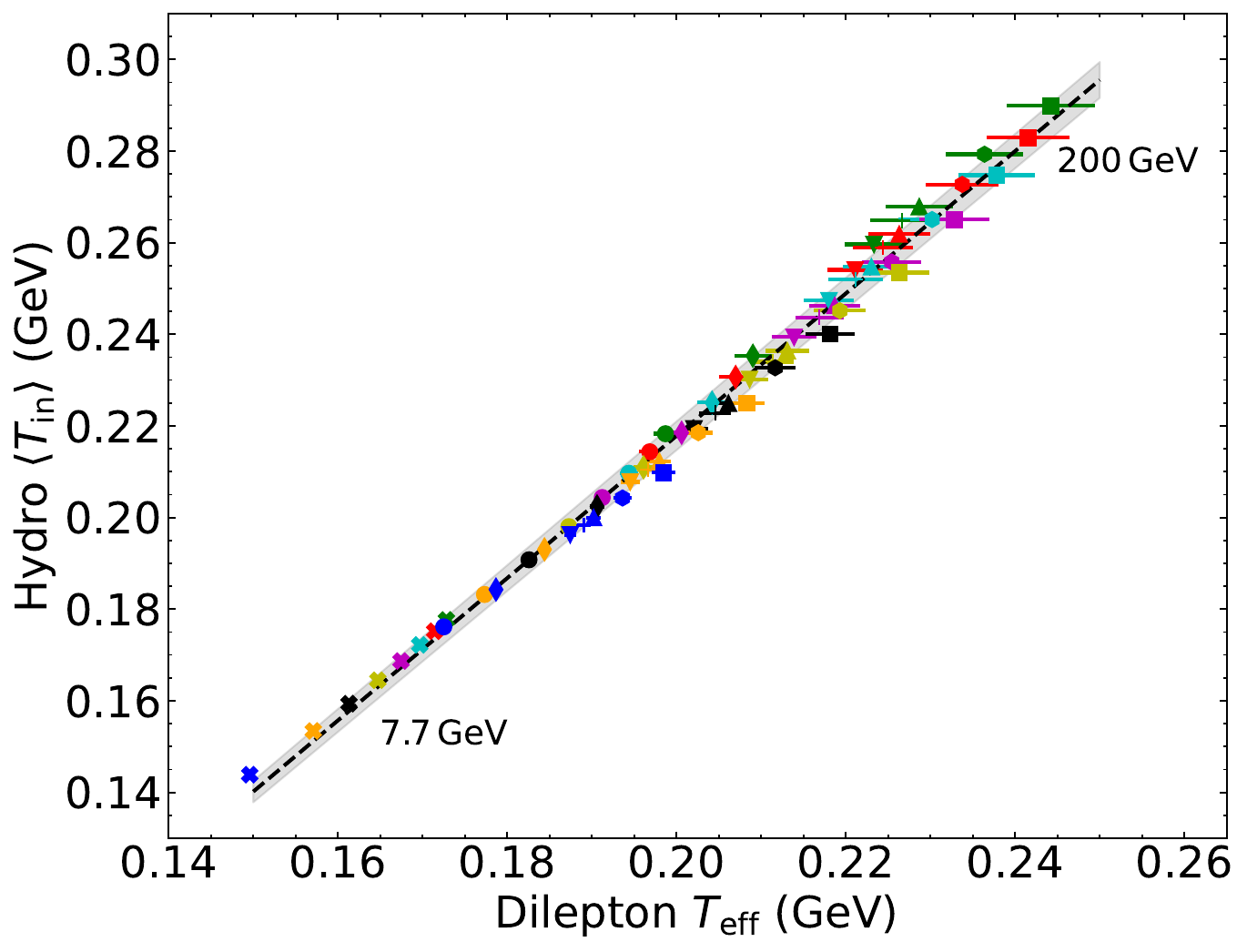}}
	\caption{The left panel shows dilepton data from the STAR collaboration (symbols) together with results of calculations described in the text (dot-dashed curves), plotted in the intermediate mass region. }
	\label{fig-3}       
\end{figure}

As a first step, and given the fact that the temperature distribution is known in the hydro fluid-dynamical cells, the accuracy of the dilepton temperature extraction as described above is estimated, and validated \cite{Churchill:2023vpt}. Next, given the multi-component approach used here which generates  hadronic phenomenology consistent with measurements \cite{Du:2022yok}, we verify whether the intermediate mass thermal dileptons produced by the theory and modeling described herein is also consistent with existing data. This is shown in the left panel of Fig. \ref{fig-3}, where the ``excess'' refers to the dilepton signal remaining after subtracting known sources, and can then be identified with a thermal contribution (references to the experimental data are to be found in \cite{Churchill:2023vpt}).  It is seen that in the mass region relevant for this study, theory and measurements agree within uncertainties. Going one step further, one may assemble the results from the effective temperature extraction for all energies and centralities considered in our study and plot them with the corresponding  ``true'' initial  temperatures extracted from the hydrodynamical simulations; this is shown in the right panel of Fig. \ref{fig-3}. One observes a striking linearity between the extracted temperature and the ``true'' temperature, reaffirming the importance of the lepton pair signal as a probe of early time dynamics. 

To conclude, we have presented results of the first derivation of  thermal dilepton production rates at finite $\mu_{\rm B}$ and at NLO. These rates have been used in a (3+1)D fluid dynamical simulation to obtain results relevant to the BES program.  Information about the QGP temperature has been shown. Note that direct information about temperature, initial or otherwise, is very rare in heavy-ion physics. Electromagnetic probes thus constitute a unique class of observables in that respect. \\

\noindent{\textbf{Acknowledgment:}}
This work was supported in part by the Natural Sciences and Engineering Research Council of Canada, in part by the U. S. Department of Energy (DOE), under grant No. DE-FG02-00ER41132, and in part by l'Agence Nationale de la Recherche (ANR), under grant ANR-22-CE31-0018. 

 \bibliography{refs}
%
%
%
%

\end{document}